\documentclass[aps,prx,amsmath,amssymb,showpacs,floatfix]{revtex4}

\usepackage{graphicx}
\usepackage{color}
\usepackage{amssymb,amsfonts,amsmath}

\usepackage{gensymb}

\newcommand{\ve}[1]{\ensuremath{\mbox{\boldmath$#1$}}}

\newcommand{\ma}[1]{\ensuremath{\mathbb{#1}}}

\newcommand{\tr}{\ensuremath{\mbox{Tr}}}

\newcommand\trans{^{\scriptstyle\mathrm T}}

\newcommand{\tauK}{\ensuremath{\tau_{\mbox{\tiny K}}}}

\DeclareMathOperator{\st}{St}
\DeclareMathOperator{\Crr}{C^{\rm rr}}

\DeclareMathOperator{\Ctr}{C^{\rm tr}}
\DeclareMathOperator{\Ctt}{C^{\rm tt}}

\newcommand{\eqnlab}[1]{\label{eq:#1}}

\newcommand{\eqnref}[1]{\eqref{eq:#1}}

\newcommand{\Eqnref}[1]{Eq.~\eqref{eq:#1}}
\newcommand{\Figref}[1]{Fig.~\ref{fig:#1}}

\newcommand{\obs}[1]{{\color{black}#1}}

\begin{document}
\title{Preferential sampling of helicity by isotropic helicoids\footnote{Postprint version of the article published on Physical Review Fluids {\bf 1} 054201 (2016) DOI: 10.1103/PhysRevFluids.1.054201}}
\author{Kristian Gustavsson}
\author{Luca Biferale}
\affiliation{Department of Physics and INFN, University of Rome ‘Tor Vergata’, 00133 Rome, Italy}

\begin{abstract}
We present a theoretical and numerical study on the motion of {\it isotropic helicoids} in complex flows. These are
particles whose motion is invariant under rotations but not under mirror reflections of the particle. This is the simplest, yet unexplored,  extension of
the much studied case of small spherical particles.
We show that heavy isotropic helicoids, due to the coupling between translational and rotational degrees of freedom, preferentially sample different helical regions in laminar or chaotic advecting flows. This opens the way to control and  engineer particles able to track complex flow structures with potential  applications to microfluidics and turbulence.
\end{abstract}

\maketitle

{\em Introduction}
Turbulent aerosols are commonly described in terms of dilute suspensions of small, heavy spherical particles in locally isotropic turbulence with motion
governed by Stokes' law \cite{Max83}.
The translational motion of a spherical particle is invariant under rotations and internal reflections of the particle. Consequently, its interaction with the flow is governed by a single parameter, the Stokes number, $\st=\tau_{\rm p}/\tauK$, defined as the ratio between the particle response time $\tau_{\rm p}$ and the characteristic time, $\tauK$, of the carrying flow.
However, in Nature and in many applications the aerosol particles are not perfect spheres.
Previous attempts to move away from the approximation of ideal spheres have been to consider
the dynamics of spheroidal particles, or of particles with even less symmetry \cite{happel2012low,guazzelli2011physical,kim2013microhydrodynamics}.
A spheroidal particle breaks rotation invariance, but it is invariant under reflections along any of its axes of symmetry.
Elongated chiral objects break both rotational and reflection invariance and they are known to develop a lateral drift even in simple shear flow. They have  been intensively studied because of their biological interest and because of the need to separate different {\it enantiomers} in micro-devices
\cite{eichhorn2010microfluidic,fu2009separation,marcos2012source,Ari13,mijalkov2013sorting,hermans2015vortex,ma2015electric,clemens2015molecular}.
Helical elongated particles with opposite chiralities at the two ends are used to track particular properties of vortex-stretching mechanisms in turbulent flows \cite{kramel2015single}.
\begin{figure}
\includegraphics[width=7cm]{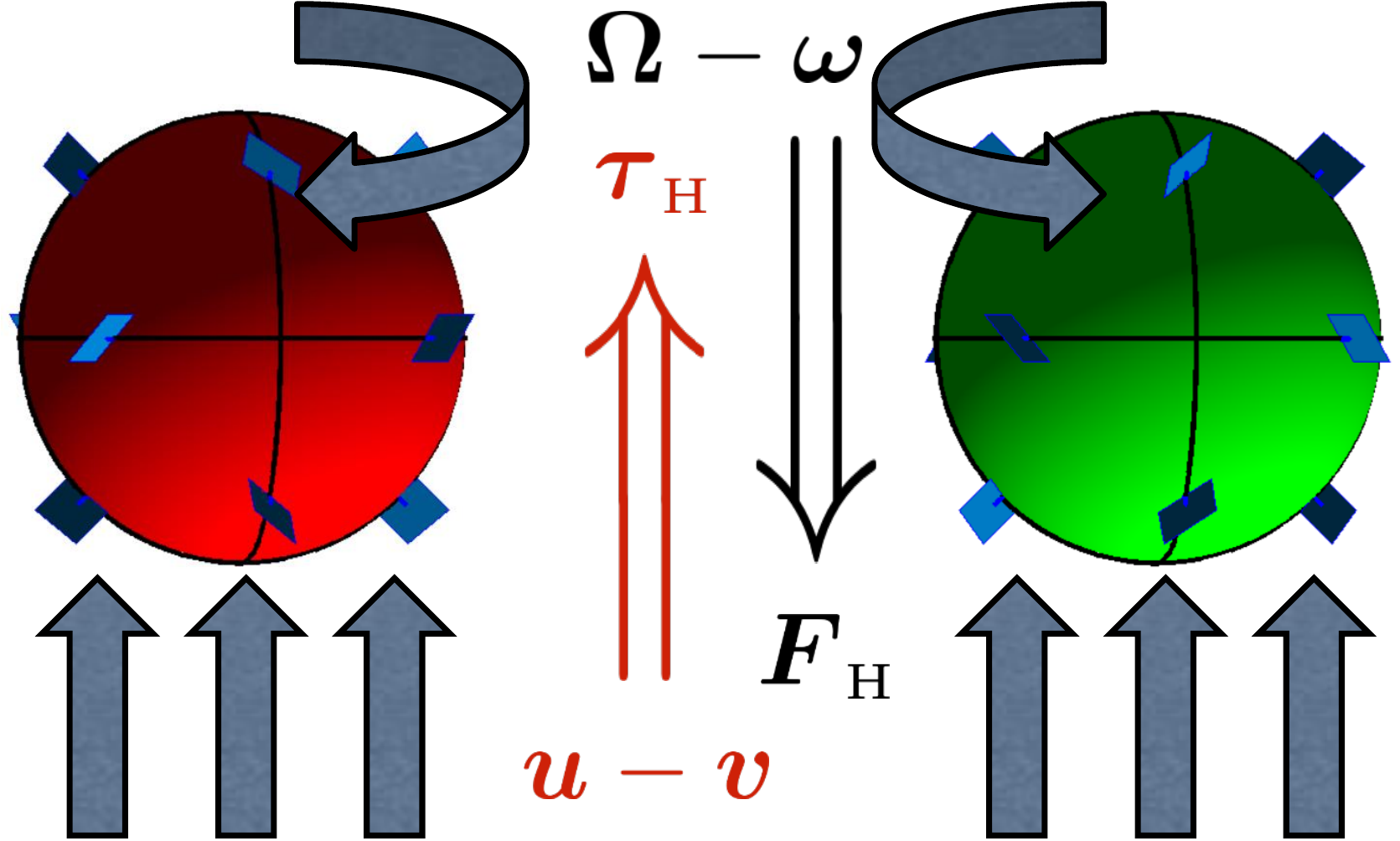}
\caption{\label{fig:Isotropic_Helicoid_Both} {\em (Color online)}. Examples of isotropic helicoids with two different chiralities: left-handed (left) and right-handed (right).
An applied velocity difference $\ve u-\ve v$ results in a torque $\ve\tau_{\rm H}$ parallel (anti-parallel) to $\ve u-\ve v$ for the right-handed (left-handed) particle. An applied difference in angular velocity $\ve\Omega-\ve\omega$ results in a force $\ve F_{\rm H}$ parallel (anti-parallel) to $\ve\Omega-\ve\omega$.
The drag force $\ve F_{\ve\rm D}$ and torque $\ve\tau_{\rm D}$ are not illustrated.
}
\end{figure}
Investigation of  the case of particles whose motion only breaks reflection invariance, while keeping rotation invariance intact, so called {\it isotropic helicoids}
\cite{Kel71,happel2012low}, is lacking. Because their internal orientation does not affect their translational motion, isotropic helicoids can be described in a lower phase-space dimension than spheroidal particles.
This makes the dynamics of isotropic helicoids {\it the simplest} extension of the much studied case of small spherical particles.
In this paper we show that isotropic helicoids may be engineered and used to find different helical regions in complex flows, with  potential new applications in microfluidics, chemistry and medical manufacturing.
The simplest observable that is statistically sensitive to spatial reflections of the carrying flow is given by its helicity averaged over a volume~$V$:
\begin{align}
\label{eq:helicity_def}
\langle H \rangle_{\rm flow} \equiv \frac{1}{V}\int_V d^3 x \,  H(\ve x,t)\,.
\end{align}
Here $H(\ve x,t)\equiv2\ve u(\ve x,t)\cdot\ve\Omega(\ve x,t)$ is the local flow helicity, defined as the product of the velocity $\ve u$ and vorticity $\ve \nabla \wedge \ve u = 2 \ve \Omega$ of the flow.
Under a spatial reflection $\ve x\to-\ve x$ of the flow, such that $H(\ve x,t)\to-H(-\ve x,t)$, the mean helicity $\langle H \rangle_{\rm flow}$ changes sign.
The average $\langle H \rangle_{\rm flow}$
 is an invariant of the three-dimensional Euler equations and  it is linked to the topology of vorticity field lines \cite{moffatt1992helicity,moffatt2014helicity}.
The correlation between energy and helicity transfer is a key open problem for many fundamental and applied turbulent flows, with  indications that intense helical regions tend to
prevent energy to flow down-scale \cite{biferale2012inverse,pouquet2010interplay,herbert2012dual}. Hence, it is  important to understand the
correlation between extreme dissipative events  and helicity in turbulence \cite{yeung2015extreme}.

In this paper we study dynamical and statistical properties of isotropic helicoids. We show that
 depending on their size, inertia and handedness,
 they have a bias to visit flow regions with certain values of flow helicity, a phenomenon  called {\it
preferential sampling}~\cite{Gus15}.
In particular, we show that, despite being heavy,  they evolve similar to {\it light} or {\it heavy} spherical particles, over- or under-sampling intense vortical structures, depending on their relative chirality with respect to the  underlying flow.

We first show that the equations of motion depend on two
 characteristic Stokes numbers, $\st_{\pm}$, related to the translational and rotational drag coefficients of the helicoid.
Finally, we analyze the dynamics of isotropic helicoids in a paradigmatic Arnlod-Beltrami-Childress (ABC) flow~\cite{dombre1986chaotic}.
We show that isotropic helicoids are indeed able to
preferentially respond to the underlying helicity.

{\em Isotropic Helicoids}
An example of an isotropic helicoid was first proposed by Lord Kelvin~\cite{Kel71}:
{\it An isotropic helicoid may be made by attaching projecting vanes to the surface of a globe in proper positions; [sic] for instance, cutting at 45\degree each, at the middles of the twelve quadrants of any three great circles dividing the globe into eight quadrantal triangles.}
\Figref{Isotropic_Helicoid_Both} shows this construction. Planar vanes are attached normal to the surface of the sphere at angles $\pm 45\degree$ to three great circles transversed clockwise. The sign of the angle determines if the particle is left-handed (left figure) or right-handed (right figure).
Each of these particles is the mirror image of the other under reflections in any plane containing the center of the particle. 
As illustrated in \Figref{Isotropic_Helicoid_Both}, if the right-handed particle experience a difference, $\ve u-\ve v$,
 between the fluid velocity $\ve u$ and its own velocity $\ve v$, such that the vector $\ve u-\ve v$ is perpendicular to the plane of one of the great circles, the particle obtains a net torque $\ve\tau_{\rm H}$ parallel to $\ve u -\ve v$ in addition to the drag force $\ve F_{\rm D}$.
This net torque is a consequence of the vanes being positioned at the same angle along the great circles.
The three great circles give three directions in which the applied velocity difference results in a torque parallel to $\ve u -\ve v$.
Linear superposition implies that a velocity difference applied in {\it any} direction results in a torque parallel to that direction.
Correspondingly, if the right-handed particle experiences a difference $\ve\Omega-\ve\omega$ between half  the
flow vorticity  and the angular velocity $\ve\omega$ of the particle, the particle obtains a force $\ve F_{\rm H}$ in the direction of $\ve \Omega -\ve \omega$
 in addition to the viscous drag torque $\ve\tau_{\rm D}$.
Similarly, the left-handed particle responds to an applied velocity (angular velocity) difference with a torque (force)
 in the opposite direction compared to the right-handed particle. The
force and torque acting on a general small and heavy isotropic helicoid in a flow are~\cite{happel2012low}
\begin{subequations}
\eqnlab{eqm_v_and_omega}
\vspace{-0.35cm}
\begin{align}
\eqnlab{eqm_v}
m\dot{\ve v}&=\ve F_{\rm D}+\ve F_{\rm H}=\Ctt(\ve u-\ve v)+\Ctr(\ve\Omega-\ve\omega)\\
I\dot{\ve \omega}&=\ve\tau_{\rm D}+\ve\tau_{\rm H}=\Crr(\ve\Omega-\ve\omega)+\Ctr(\ve u-\ve v)\,.
\eqnlab{eqm_omega}
\end{align}
\end{subequations}
Here $\ve u$ and $\ve\Omega$ are evaluated at the position $\ve r_t$ of the particle at time $t$, overdots denote time derivatives, and $m$ and $I$ are the mass and moment-of inertia of the particle.
Due to the symmetries of isotropic helicoids the {moment-of inertia}, translation, coupling, and rotation tensors are given by scalar quantities {$I_0$,} $\Ctt$, $\Ctr$, and $\Crr$ times the identity matrix.
In order for the system to dissipate energy, preventing {exponentially growing} solutions, the following condition must be fulfilled~\cite{happel2012low}:
$(C^{{\rm tr}})^2<C^{{\rm tt}}C^{\rm rr}$.
The coefficients $\Ctt$ and $\Crr$ are positive while $\Ctr$ can take either sign, corresponding to a reflection-invariant ($\Ctr=0$), left-handed ($\Ctr<0$), or right-handed ($\Ctr>0$) particle. When $\Ctr=0$, Eqs.~\eqnref{eqm_v_and_omega} reduce to Stokes' drag force and torque.
When $\Ctr\ne 0$ the force $\ve F_{\rm H}$ and torque $\ve\tau_{\rm H}$ illustrated in \Figref{Isotropic_Helicoid_Both} couples the translational and
rotational dynamics.
This coupling breaks invariance of the motion under mirror reflections of the particle.
Consider for example a flow region where one component of $\Omega_i$ is very large compared to $\omega_i$, and to {$(u_i-v_i)\Ctt/\Ctr$}.
In this region, \Eqnref{eqm_v} may be approximated by $m\dot{v}_i\approx\Ctr\Omega_i$.
Depending on the relative sign of $\Ctr$ and $\Omega_i$ the particle accelerates either along the vorticity component, or opposite to it.
Large vortices may thus accelerate particles to different regions in the flow depending on their helicity.
This example shows that isotropic helicoids, in contrast to spherical particles,  distinguish flow structures of different parity.
Reflection invariance in the dynamics of isotropic helicoids may be broken in two ways: by the flow itself, statistically or locally, or by the dynamics as illustrated in the example above.
One example is homogeneous isotropic turbulence where spatial reflection symmetry of the flow is often assumed in a statistical sense,
but locally parity-breaking structures do form.

{\em Dimensionless control parameters.}
We characterize small-scale fluctuations of the flow by the characteristic Eulerian speed and length scales: $u_0$ and $\eta_0$, respectively, and by the Lagrangian time scale $\tauK\equiv 1/\sqrt{\tr\langle\ma A\ma A\trans\rangle}\sim\eta_0/u_0$, where $\ma A$ is the fluid gradient matrix with components $A_{ij}\equiv\partial_ju_i$ and the angular brackets denote a spatial average as in Eq. (1).
Together with the dimensional parameters governing the dynamics in Eqs~\eqnref{eqm_v_and_omega}: $\Ctt$, $\Crr$, $\Ctr$, $m$, and $I$
it is possible to form four independent dimensionless parameters.
These are the {\it Stokes number}, $\st\equiv m/(\tauK\Ctt)$, the {\it structural number}, $S\equiv 3m\Crr/(10I\Ctt)$,
the {\it helicoidal number}, $C_0\equiv9\sqrt{m/(10I)}\Ctr/\Ctt$ and the {\it dimensionless radius},  $\overline{a}\equiv\sqrt{5I/(2m\eta_0^2)}$.
The parameters $\st$ and $\st_{\rm r}\equiv 3\st/(10S)$ determine the translational and rotational inertia of the particle when $C_0=0$. Here 
 $S$ is proportional to the ratio of
these quantities such that a spherical particle has $S=1$.
In addition, $C_0$ inherits the properties of $\Ctr$: its sign determines the chirality of the particle and the condition to prevent 
exponentially growing  solutions becomes $|C_0|<\sqrt{27S}$.
When $C_0=0$, the motion simplifies to that of an isotropic particle, and if further $S=1$ the motion is that of a spherical particle with Stokes number $\st$,  rotational Stokes number $\st_{\rm r}=3\st/10$ and dimensionless radius $\overline{a}$.
For general values of $C_0$ the parameters $\overline{a}$ and $S$ can take any positive values. For helicoids resembling spheres we interpret $\overline{a}$ as an estimate of the particle size in units of the Eulerian length scale $\eta_0$.
To simplify the discussion in what follows, we assume $S=1$ (see Appendix A  for expressions with general $S$). This value corresponds to global shapes close to that of a sphere (see \Figref{Isotropic_Helicoid_Both}).
Using the dimensionless variables $t'=t/\tauK$, $\ve r'=\ve r/\eta_0$, $\ve v'=\ve v\tauK/\eta_0$, and $\ve\omega'=\ve\omega\tauK$
(primes will be dropped from here on) the equations of motion corresponding to the force and torque in Eqs.~\eqnref{eqm_v_and_omega} become $\dot{r}_i=v_i$ for the position components and:
\begin{equation}
\begin{pmatrix}
\dot{v}_i\cr
\dot{\omega}_i
\end{pmatrix}
=\ma D
\begin{pmatrix}
u_i-v_i\cr
\Omega_i-\omega_i
\end{pmatrix}
\,,\hspace{0.5cm}
\ma D=
\frac{1}{\st}\begin{pmatrix}
1 & \frac{2\overline{a} C_0}{9} \cr
\frac{5 C_0}{9\overline{a}} & \frac{10}{3}
\end{pmatrix}
\,
\eqnlab{eqm_vomega_rescaled}
\end{equation}
for the components of velocity and angular velocity.
Due to rotational invariance the matrix $\ma D$ does not mix different Cartesian components.
\begin{figure*}
\includegraphics[width=16cm]{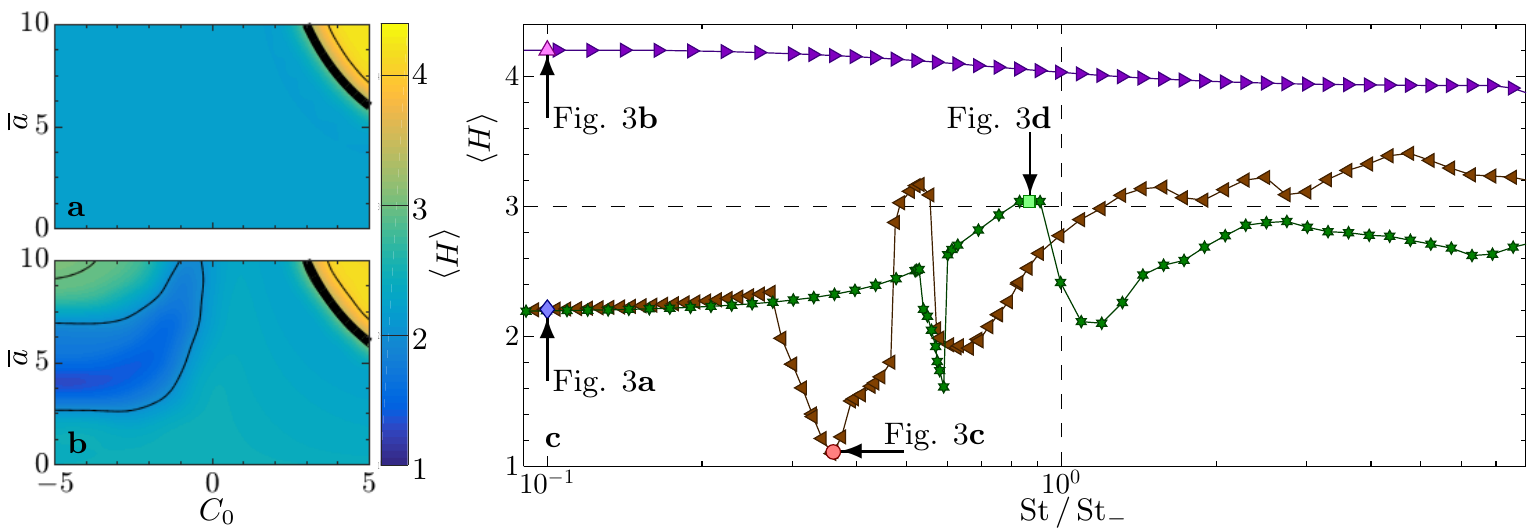}
\caption{\label{fig:ABC2} {\em (Color online)}. Average helicity, $\langle H \rangle$, for different parameters. 
Shown is the contour plot against $C_0$ and $\overline{a}$ with $S=1$ and $\st=0.1\st_-$ ({\bf a}) and $\st=0.5\st_-$ ({\bf b}).
The average is obtained from numerical simulations of Eqs.~\eqnref{eqm_vomega_rescaled} and \eqnref{u_ABC}. 
Black thick lines show the predicted transition (\ref{eq:transition}).
{\bf c}: average helicity  $\langle H \rangle$ against $\st/\st_-$ for different helicoidal numbers: $C_0=-5$ (brown left-pointing triangle), $C_0=0$ (green star), $C_0=5$ (purple right-pointing triangle) {, $\overline{a}=10$ and $S=1$}. Vertical dashed line shows $\st=\st_-$. Mean flow helicity \obs{$\langle H \rangle_{\rm flow}=3$.} (horizontal dashed line)}
\end{figure*}

\begin{figure*}
\includegraphics[width=16cm]{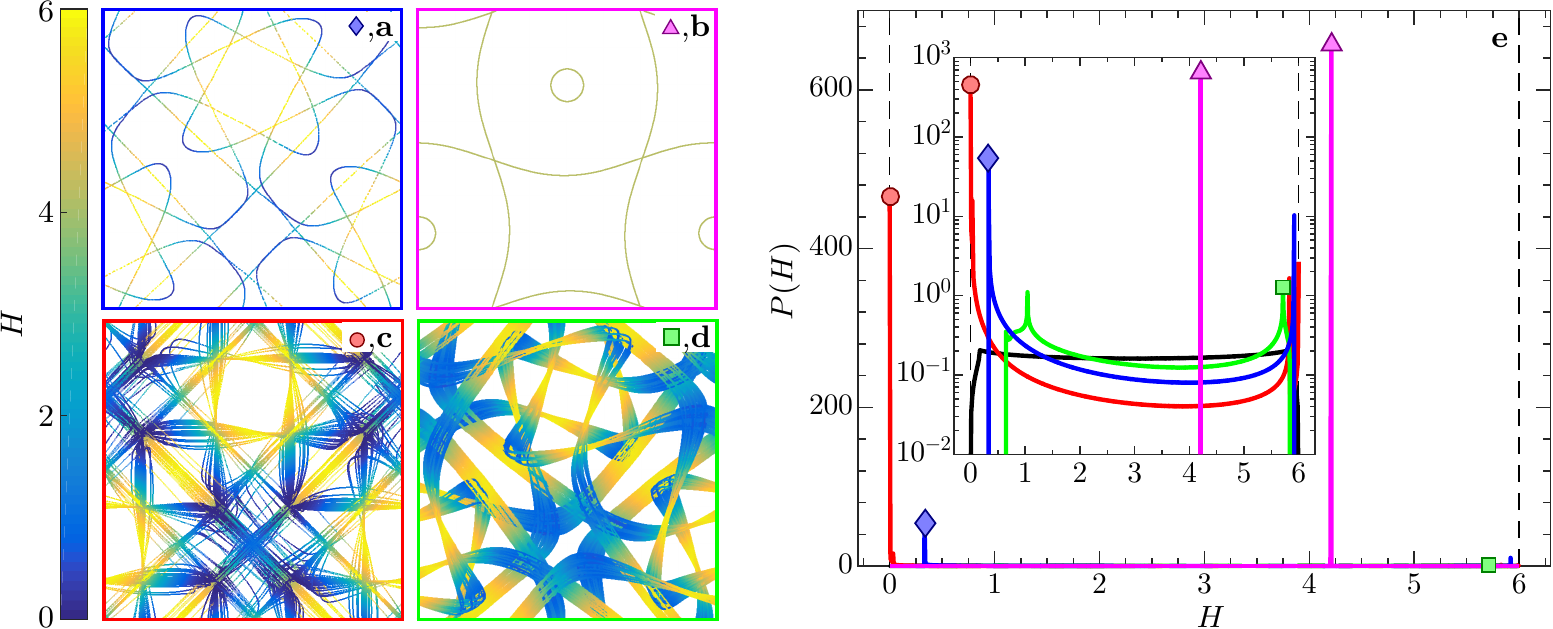}
\caption{\label{fig:ABC3} {\em (Color online)}
{\bf a}--{\bf d} {Each panel shows $100$} trajectories {for 100 time units,after an initial transient. The trajectories are} projected on the $x$-$y$ plane (ranges from {$0$ to $2\pi$}).
Trajectory positions are colored depending on the local helicity of the underlying flow.
The parameters  are $\overline{a}=10$ and $S=1$, and  ({\bf a})  $C_0=-5$, $\st=0.1\st_-$,  ({\bf b}) $C_0=5$, $\st=0.1\st_-$,  ({\bf c})
 $C_0=-5$, $\st=0.36\st_-$, ({\bf d}) $C_0=0$, $\st=0.87\st_-$.
 The corresponding parameter values are also indicated in Fig. (\ref{fig:ABC2}){\bf c}.
({\bf e}):  PDF of helicity, $P(H)$, for the trajectories shown in {\bf a}--{\bf d} and for the ABC flow (black line). \obs{The bin width is  $10^{-4}$,  
markers correspond to the most probable value [same symbols as in  panels {\bf a}-{\bf d}]. 
The inset shows data on a  log-lin scale.}
}
\end{figure*}
It is instructive to estimate the degree of compressibility
experienced by particles with small $\st$.
From \eqnref{eqm_vomega_rescaled} we find:
\begin{equation}
\label{eq:compressibility}
\ve\nabla\cdot\ve v \sim -\,\frac{\st}{27-C_0^2}\!\Big(\!27 \tr[\ma A^2] -\frac{9 \overline{a} C_0}{5} \tr[\ma A \ma V]\!\Big)\!+o(\st)\,,
\end{equation}
where we have introduced the matrix $\ma V$ with components $V_{ij}\equiv\partial\Omega_i/\partial r^j$. Equation 
\Eqnref{compressibility} shows that helicoids experience different flow topology depending
on the relative magnitude of $\tr \ma A^2$ and $\tr \ma A \ma V$. In particular, in regions of strong, say positive,
local helicity we typically have $\ve \Omega \sim k \, \ve u$  leading to an estimate for the compressibility:
\begin{align}
\label{eq:compressibility_aligned}
\ve\nabla\cdot\ve v  \propto -\tr[ \ma A^2] [ 27 - 9  \overline{a} C_0k/5]\,.
\end{align}
Depending on the values of $k, \overline{a}$ and $C_0$, \Eqnref{compressibility_aligned} indicates the tendency to escape or to be trapped in regions where $\tr [\ma A^2]$ is positive or negative.
The decomposition $\tr [\ma A^2]=\tr[\ma S^2] -2\ve\Omega^2$, with $\ma S\equiv (\ma A+\ma A^{\rm T})/2$ shows that helicoids may be attracted either to regions of strong vorticity ($\tr[\ma S^2]\ll 2\ve\Omega^2$) similar to light spherical particles, or to regions of strong shear ($\tr[\ma S^2]\gg 2\ve\Omega^2$) similar to heavy spherical particles.
Which behaviour is chosen depends on the sign of the local flow helicity and on the parameters of the helicoid.

In order to better understand the dynamics of \eqnref{eqm_vomega_rescaled}, we diagonalize it (see Appendix A).
It turns out that the evolution is characterized by the
two eigenvalues $\st_{\pm}/\st$ of $\ma D$:
\begin{align}
\st_\pm=\Big(39\pm\sqrt{40C_0^2+441}\,\Big)/18\,.
\eqnlab{Stpm}
\end{align}
These define two important inertial scales for the isotropic helicoid.
When $C_0=0$,  $\st_\pm$ equal the translational and rotational drag coefficients of an isotropic particle. When $C_0^2$ is close to its limiting value 27, we find that $\st_-$ approaches zero while $\st_+$ takes a finite value $13/3$.
This separation of scales, $\st_+/\st_-\to\infty$ as $C_0^2\to 27$, allows for rich dynamics, with different non-trivial behavior when $\st\sim\st_-$ or when $\st\sim\st_+$. 
In particular, we expect a high sensitivity of the helicoids trajectories on the helicity of the underlying flow. 
In what follows, we consider the dynamics of  helicoids driven by a model helical flow: the well-studied case of an ABC flow~\cite{dombre1986chaotic,aref1990chaotic}.


{\em ABC flow}
Being a solution of the three-dimensional Euler equations with chaotic streamlines, the ABC flow has been the subject of many studies in turbulence theory.
Ref.~\cite{moffatt2014helicity} argues that the Euler equations has steady solutions consisting of  patches of ABC flows
 connected by vortex sheets.
%
The Eulerian velocity field of the ABC flow with equal coefficients is given by (in dimensionless variables)
\begin{align}
\eqnlab{u_ABC}
\ve u=(\cos y+\sin z,\cos z+\sin x,\cos x+\sin y)\,.
\end{align}
Although being time-independent, fluid particles governed by Eq.~\eqnref{u_ABC} show chaotic behavior
\cite{dombre1986chaotic,wang1991quantification}, with dimensionless time scale $\tauK=1/\sqrt{3}$.
The ABC flow has the property that $\ve\Omega=k\ve u$ with $k=1/2$ at all positions. This implies that only structures with non-negative local helicity, $H(\ve x) \ge 0$, are encountered in the flow.
The local particle compressibility is given by \Eqnref{compressibility_aligned} with $k=1/2$ if $\st$ is small enough.
Moreover, for the ABC-flow $\tr \ma A^2$ and $H$ are related by $\tr[\ma A^2] = \langle H \rangle_{\rm flow} -H$\obs{, where $\langle H\rangle_{\rm flow}=3$ according to Eqs.~\eqnref{helicity_def} and \eqnref{u_ABC}.}
From \Eqnref{compressibility_aligned} it follows that there exists a critical combination
\begin{equation}
\label{eq:transition}
\overline{a}C_0 = 30
\end{equation}
which distinguishes whether particles are attracted to regions with high or
low values of $H$.
\obs{This phase transition is illustrated in \Figref{ABC2}{\bf a}.
The average helicity, $\langle H \rangle$,
 evaluated along particle trajectories is plotted against $C_0$ and $\overline{a}$ with $\st/\st_-=0.1$.
When $\overline{a}$ is larger than its critical value the 
helicoids are able to oversample the underlying helicity by forming
 stable periodic orbits with almost constant helicity.
It is important to stress that helicoids with a certain handedness  are attracted only 
by vortices with that chirality (as opposed to light spherical particles which do 
not distinguish the vortex helicity). On the other hand, 
below the transition the helicoids are attracted by straining regions.
For larger values of $\st/\st_-$, the dynamics becomes richer, as shown in \Figref{ABC2}{\bf b} for $\st/\st_-=0.5$, 
but still  particles oversample (undersample) the flow helicity above (below) the critical line (\ref{eq:transition}). 

The transition in the preferential sampling of $H$, as a function of the Stokes number
is quantified  in \Figref{ABC2}{\bf c}, where we show
 $\langle H\rangle$  for left- ($C_0=-5$), neutral- ($C_0=0$), and right- ($C_0=5$) handed helicoids.  
The general trend is complicated as expected for most low-dimensional chaotic advection problems. Nevertheless, in some 
limits their dynamics develop a more systematic behaviour.  
For large values of $\st/\st_+$ the mean helicities in \Figref{ABC2}{\bf c} approach
 that of the flow for all values of $C_0$ (not shown). For $\st/\st_-$ much smaller than unity,  the mean helicity 
collapses to the two  phases predicted by the transition (\ref{eq:transition}).
Corresponding trajectories are 
shown in \Figref{ABC3}{\bf a},{\bf b}. In both cases we observe stable periodic orbits with higher (\Figref{ABC3}{\bf b}) or
lower (\Figref{ABC3}{\bf a}) mean helicity compared to the underlying flow.
We remark that in this limit the preferential sampling of helicity is extremely focused, especially for the case shown in \Figref{ABC3}{\bf b}, leading to strong peaks in the corresponding probability distribution functions, as shown in \Figref{ABC3}{\bf e}. Concerning the region where $\st/\st_- \sim O(1)$, right-handed particles continue to follow stable periodic orbits (see also Fig. \ref{fig:stpmA} in  Appendix A), while 
left-handed particles tend to be scattered to larger spatial regions (\Figref{ABC3}{\bf c}). 
Still, as shown in Fig.  3({\bf e}) left-handed helicoids may show significant bias toward different values of helicity
compared to right-handed helicoids. Finally, neutral particles are generally 
more scattered  (\Figref{ABC3}{\bf d}) and their helicity distribution is
closer to the one of the flow [green and black lines in \Figref{ABC3}{\bf e}, respectively]. 
}


We remark that \Eqnref{eqm_vomega_rescaled} is derived assuming that the particles are {\it small} compared to the smallest length scale $\eta_0$ of the flow. To enhance the  effects of chirality it is important that the normalized size, $\overline{a}$, is larger than unity. One example of how to satisfy these constraints is to attach to the particle in \Figref{Isotropic_Helicoid_Both}
four  small and heavy satellite particles at the corners of a tetrahedron, connected  by  thin rods of length larger than~$\eta$. If the satellites are small enough, they will not contribute to the particle-fluid interaction but they may give a substantial contribution to the moment of inertia, leading to $\overline{a} >1$ without contradicting the requirements for the validity of the equations of motion.\\

{\em Conclusions.}
We have presented an analysis of dynamical and statistical properties of small, heavy, dilute isotropic helicoids in chaotic flows.
Their motion constitutes the simplest generalization to that of small spherical particles. We have shown that their dynamics is ruled by two distinct characteristic time scales, $\st_\pm$.  
\obs{
Chiral particles are often encountered in bio-fluidic systems \cite{eichhorn2010microfluidic,fu2009separation,marcos2012source,Ari13,mijalkov2013sorting,hermans2015vortex,ma2015electric,clemens2015molecular,Ruiz04}.
 It is also known that a liquid made of chiral molecules (or a suspension of chiral molecules)  is described by an extended version
 of the Navier-Stokes equations, with additional stresses forbidden for non chiral components \cite{Andreev2010}.   Here, we have shown that 
isotropic helicoids can be used in direct numerical simulations and/or in laboratory experiments as {\it smart} probes,  able to track dynamically relevant topological information of the carrying flow.}
An important parameter  that we did not discuss is the structural number $S$.
In this paper the case $S=1$ was considered.
Formulae for general values of $S$ are qiven in  Appendix A.
It turns out that the chiral symmetry-breaking term in Eq.(\ref{eq:compressibility}), $\tr[\ma A \ma V]$,
 becomes more important as $S$ decreases.
How to construct particles with given parameters, $\st$, $S$, $\overline{a}$, and $C_0$ is an open theoretical and experimental challenge.
{Let us also note that other forces might be important, e.g. gravity or the history force \cite{Olivieri2014}, 
  depending on the density ratio between the helicoids and the surrounding flow,  the combined effects of all of them is a key problem that should be
 addressed in future work.}
We acknowledge funding from the European Research Council under the
European Union's Seventh Framework Programme, ERC Grant Agreement NewTURB No 339032.


\section{Appendix A}
\subsection{Equation of motion for isotropic helicoids}
\label{appendix:Eqm}
In terms of the dimensionless variables in Sec. III the equations of motion {become}
\begin{equation}
\begin{pmatrix}
\dot{v}_i\cr
\dot{\omega}_i
\end{pmatrix}
=\ma D
\begin{pmatrix}
u_i-v_i\cr
\Omega_i-\omega_i
\end{pmatrix}
\,,\hspace{0.5cm}
\ma D=
\frac{1}{\st}\begin{pmatrix}
1 & \frac{2\overline{a} C_0}{9} \cr
\frac{5 C_0}{9\overline{a}} & \frac{10S}{3}
\end{pmatrix}
\,
\eqnlab{eqm_vomega_rescaled_appendix}
\end{equation}
for each component $i=1,2,3$ of $\ve v$ and $\ve\omega$ [when $S=1$ \Eqnref{eqm_vomega_rescaled_appendix} simplifies to Eq.~(3)].

\begin{figure}
\includegraphics[width=8cm]{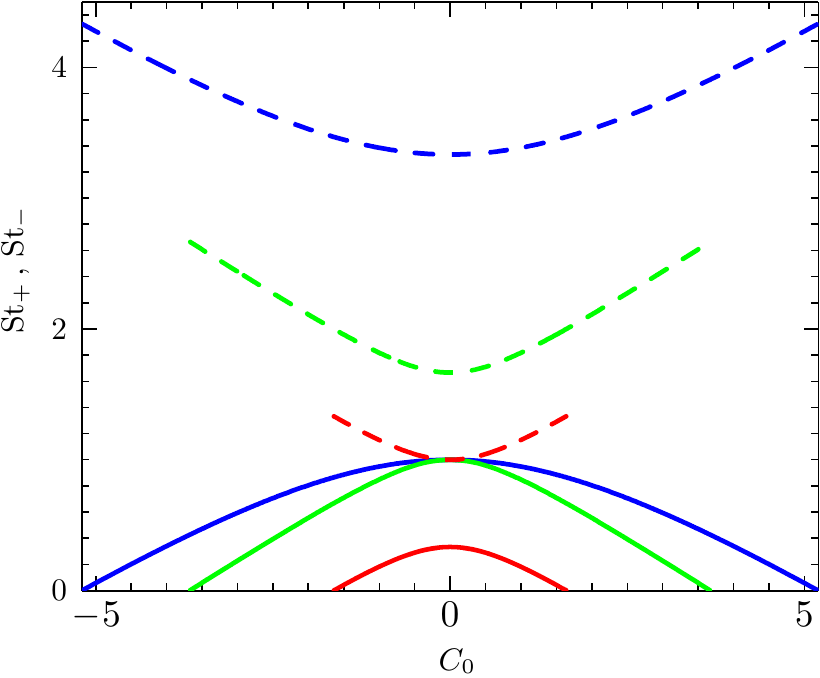}
\caption{\label{fig:stpmB}{\em (Color online)} Plot of $\st_+$ (dashed lines) and $\st_-$ (solid lines) in \Eqnref{Stpm_appendix} as a function of $C_0$ for different values of $S$: $S=0.1$ (red), $S=0.5$ (green), and $S=1$ (blue).
}
\end{figure}

Diagonalisation of the matrix $\ma D$ gives two eigenvalues $d_\pm$ and two normed eigenvectors $\xi_\pm$
\begin{subequations}
\begin{equation}
d_\pm=\frac{1}{18\st}\left(9+30S\pm\sqrt{40C_0^2+9(3-10S)^2}\right)
\end{equation}
\begin{equation}
\ve\xi_\pm\!\!=\!
\frac{1}{\sqrt{100C_0^2\!+\!36\overline{a}^2(3d_\pm\st-10S)^2}}
\begin{pmatrix}
6 \overline{a}(3d_\pm\!\st\! - 10 S)\! \cr
10C_0
\end{pmatrix}\,.
\end{equation}
\end{subequations}
We define the matrices
\begin{align}
\ma C&=
\begin{pmatrix}
d_- & 0 \cr
0 & d_+
\end{pmatrix}
\,\mbox{ and }\,
\ma X=
\begin{pmatrix}
1 & 1 \cr
\xi_{-,2}/\xi_{-,1} & \xi_{+,2}/\xi_{+,1}
\end{pmatrix}
\end{align}
such that
\begin{align}
\ma X^{-1}&=
\frac{1}{3\st(d_+-d_-)}
\begin{pmatrix}
10 S - 3 d_-\st  & -\frac{2C_0\overline{a}}{3} \cr
3 d_+\st - 10 S & \frac{2C_0\overline{a}}{3}
\end{pmatrix}
\end{align}
and insert $\ma D=\ma X\ma C\ma X^{-1}$ in the equation of motion \eqnref{eqm_vomega_rescaled_appendix}
\begin{equation}
\ma X^{-1}\begin{pmatrix}
\dot{v}_i\cr
\dot{\omega}_i
\end{pmatrix}
=\ma C\ma X^{-1}
\begin{pmatrix}
u_i-v_i\cr
\Omega_i-\omega_i
\end{pmatrix}\,.
\end{equation}
For each component $i$ we introduce the variables
\begin{align}
\begin{pmatrix}
\zeta_{-,i}\cr
\zeta_{+,i}
\end{pmatrix}
\equiv
\ma X^{-1}
\begin{pmatrix}
v_i\cr
\omega_i
\end{pmatrix}
\end{align}
and the fields
\begin{align}
\begin{pmatrix}
u_{-,i}\cr
u_{+,i}
\end{pmatrix}
\equiv
\ma X^{-1}
\begin{pmatrix}
u_i\cr
\Omega_i
\end{pmatrix}
\end{align}
to obtain the diagonalised equations of motion
\begin{subequations}
\begin{equation}
\dot{\zeta}_{-,i}=\frac{\st_-}{\st}(u_{-,i}-\zeta_{-,i})
\end{equation}
\begin{equation}
\dot{\zeta}_{+,i}=\frac{\st_+}{\st}(u_{+,i}-\zeta_{+,i})\,.
\end{equation}
\eqnlab{eqm_diagonal}
\end{subequations}
Here we have defined two characteristic Stokes numbers $\st_\pm\equiv d_\pm\st$ of the dynamics (Eq.~(6) in the main paper when $S=1$, see also \Figref{stpmB}):
\begin{align}
\st_\pm=\frac{1}{18}\left(9+30S\pm\sqrt{40C_0^2+9(3-10S)^2}\right)\,.
\eqnlab{Stpm_appendix}
\end{align}
Eqs.~\eqnref{eqm_diagonal} are implicitly coupled through the position dependence in $u_{\pm,i}=u_{\pm,i}(\ve r_t,t)$, where $\ve r_t$ is the solution of $\dot{r}_i=v_i=\zeta_{-,i}+\zeta_{+,i}$.
We remark that the characteristic Stokes numbers $\st_\pm$ may be well separated in the sense that the relative magnitude $\st_+/\st_-$ approaches infinity as $C_0$ approaches its limiting values $\pm\sqrt{27S}$.

\subsection{Stable periodic trajectories in ABC flow, $\st/\st_- \ll 1$}
\begin{figure}
\includegraphics[width=8cm]{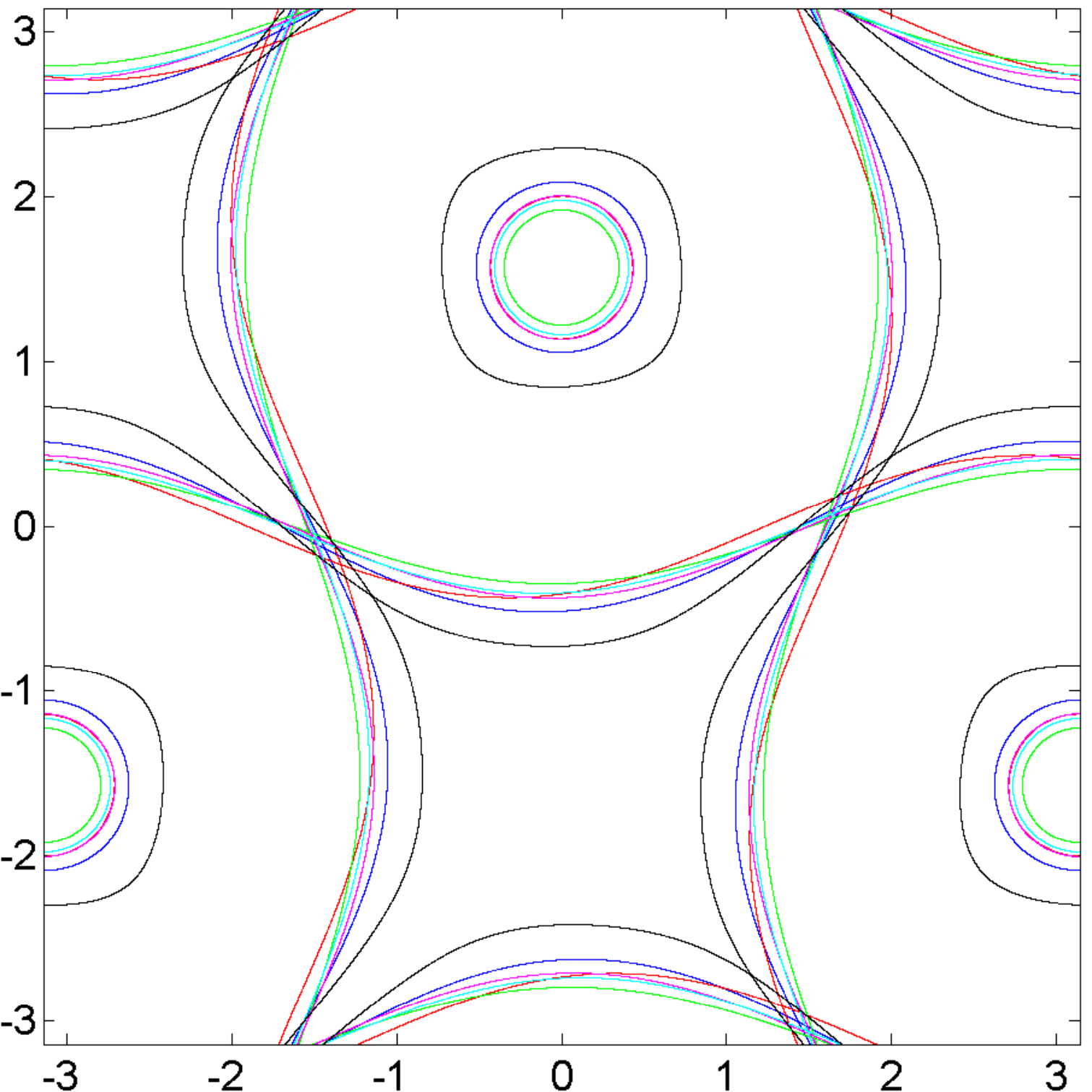}
\caption{\label{fig:stpmA}{\em (Color online)} Examples of stable periodic solutions with approximately constant helicity plotted for a number of parameters.
Curves are plotted in the order \{blue, red, green, magenta, cyan,and  black\}
with the mean and variance of helicity $\langle H\rangle=\{3.74,4.02,4.20,3.97,4.05,3.06\}$, and
 $\langle H^2\rangle-\langle H\rangle^2=\{14.5,3.87,0.77,3.53,2.32,72.6\}\cdot 10^{-5}$.
The parameters are $C_0=\{1.64,1.64,5,5,5,1.64\}$, $\st/\st_-=\{1,0.1,0.1,1,0.5,1\}$, $\overline{a}=\{10,30,10,10,10,30\}$, and 
$S=\{0.1,0.1,1,1,1,0.1\}$.
}
\end{figure}

As discussed in Sec. IV, isotropic helicoids of the same handedness of the underlying ABC flow possess stable periodic trajectories that oversample the flow helicity  for
\begin{equation}
\label{eq:transition2}
\overline{a}C_0 > 30 S
\end{equation}
 [which leads to Eq. (8)  for $S=1$] and $\st/\st_-$ small enough. These periodic orbits are approximate isolines of helicity with a value that can be tuned by changing the  structural properties of the particles. For example, in Fig. (\ref{fig:stpmA}) we show some of the 
 trajectories that one can get at changing $C_0,S, \overline{a}$ and $\st/\st_-$.
On the other hand, below the transition: 
\begin{equation}
\label{eq:transition2A}
\overline{a}C_0 < 30 S
\end{equation}
the helicoids have periodic orbits that undersample the flow helicity with a broader distribution. 



\end{document}